\documentclass[preprint,showpacs,amsmath,amssymb,aps,prd]{revtex4}

\newcommand{\kompleksni}{\ensuremath{\mathbb{C}}}

\newcommand{\prirodni}{\ensuremath{\mathbb{N}}}

\begin{document}

\title{A finiteness bound for the EPRL/FK spin foam model}

\author{Aleksandar Mikovi\' c}
 \email{amikovic@ulusofona.pt}
\altaffiliation[Member of ]{Grupo de F\'isica Matem\'atica da
Universidade
de Lisboa,
Av. Prof. Gama Pinto, 2, 1649-003 Lisboa, Portugal}
\affiliation{Departamento de Matem\'atica, Universidade Lus\'{o}fona de
Humanidades e Tecnologia,
Av. do Campo Grande, 376, 1749-024, Lisboa, Portugal}
\author{Marko Vojinovi\'c}
 \email{vmarko@cii.fc.ul.pt}
\affiliation{Grupo de F\'isica Matem\'atica da Universidade de Lisboa,
Av. Prof. Gama Pinto, 2, 1649-003 Lisboa, Portugal}

\date{\today}

\begin{abstract}
We show that the EPRL/FK spin foam model of quantum gravity has an absolutely convergent partition function if the vertex amplitude is divided by an appropriate power $p$ of the product of
dimensions of the vertex spins. This power is independent of the spin foam 2-complex and
we find that $p> 2$ insures the  convergence of the state sum. Determining the convergence of the state sum for the values $0\le p\le 2$ requires the knowledge of the large-spin asymptotics of the vertex amplitude in the cases when some of the vertex spins are large and other are small. 
\end{abstract}

\pacs{04.60.Pp}

\maketitle

\section{\label{IntroductionSection}Introduction}

Spin foam models are quantum gravity theories where the quantum geometry of
spacetime is described by a colored two-complex where the colors are
the spins, i.e. the irreducible $SU(2)$ group representations and the corresponding
intertwiners. By assigning appropriate weights for the simplexes of the
2-complex and by summing over the spins and the intertwiners, one obtains a
state sum that can be interpreted as the transition amplitude for the boundary
quantum geometries, which are described by spin networks \cite{lqg}. A spin foam
state sum can be considered as a path integral for general relativity.

The most advanced spin foam model constructed so far is the EPRL/FK model,
introduced in \cite{elpr,fk}. The finiteness, as well as the semiclassical
properties of a spin foam model, depend on the large-spin asymptotics of the
vertex amplitude. This asymptotics was studied  in
\cite{ConradyFreidel,Barret2009eeprl,Barret2009leprl} for the EPRL/FK case. The study of the finiteness of the model was started in
\cite{Perini2008}, where only two simple spin foam amplitudes were studied
(equivalent to loop Feynman diagrams with 2 and 5 vertices) in the Euclidean
case. It was concluded that the degree of divergence of these two spin foam
transition amplitudes depends on a choice of the normalization of the vertex
amplitude. This normalization is a power of the product of the dimensions of the
spins and the intertwiners which label the faces and the edges of the 4-simplex dual
to a spin-foam vertex. 

One can exploit this freedom in the definition of the EPRL/FK vertex amplitude in order to
achieve the finiteness of the model. Namely, an EPRL/FK vertex amplitude can be introduced such that
it is the original one divided by a positive power $p$ of the
product of dimensions of the vertex spins $\Delta_v$. This new amplitude will give the state sum with better convergence properties, and one can try to find a range of $p$ for which the state sum is convergent. In this paper we will show that there are such values of $p$ which are independent from the spin-foam 2-complex.

Note that an equivalent approach was used in the case of the Barrett-Crane spin foam
model, where the finiteness was achieved by introducing an appropriate edge
amplitude \cite{cpr, bcht}. This is an equivalent approach to our approach because a state sum with a dual edge amplitude $A_3(j)= (\dim\,j_1 ... \dim\,j_4)^q$ and a vertex amplitude $A_4(j)$ is the same as the state sum with $\tilde A_3 (j) = 1$  and $\tilde A_4 (j) = (\Delta_v)^p (j)\, A_4 (j)$, where $p$ is an appropriate power.

Our paper is organized such that in section \ref{VertexSection} we describe the EPRL/FK spin foam model and discuss the large-spin asymptotic properties of the vertex amplitude. In section \ref{sect3} we show that the vertex amplitude divided by the product of the dimensions of the vertex spins is a bounded function of the spins. In section \ref{FinitenessSection} we introduce a rescaled EPRL/FK vertex amplitude, which is the original amplitude divided by the product of the dimensions of the vertex spins raised to a power $p$. We prove that the corresponding state sum is absolutely convergent for $p>2$ by using the amplitude estimate from section \ref{sect3}. In section \ref{ConclusionSection} we discuss our results and present conclusions.

\section{\label{VertexSection}The vertex amplitude}

The EPRL/FK spin foam model state sum  is given by
\begin{equation} \label{StateSumaIntertw}
Z(T) = \sum_{j,\iota } \prod_{f\in T^*} A_2 (j_f) \prod_{v\in T^*}
W (j_{f(v)},\iota_{e(v)}) \,,
\end{equation}
where $T$ is a triangulation of the spacetime manifold, $T^*$ is the dual simplicial complex, while $e$, $f$ and $v$ denote the edges, the faces and
the vertices of $T^*$, respectively. The sum in (\ref{StateSumaIntertw}) is over all possible assignements of $SU(2)$ spins $j_f$ to the faces of $T^*$ (triangles of $T$) and over the corresponding intertwiner assignemets $\iota_e$ to the edges of $T^*$ (tetrahedrons of $T$). $A_2$ is the face amplitude, and it can be fixed
to be
\begin{equation}
A_2 (j) = \textrm{dim}\, j = 2j + 1
\,,\label{fa}
\end{equation} 
by using the consistent glueing reguirements for the transition amplitudes between three-dimensional boundaries, see \cite{Bianchi2010}.

The vertex amplitude $W$ can be written as
\begin{equation}
W (j_{f}, \iota_e ) = \sum_{k_e \ge 0} \int_0^{+\infty} d\rho_e (k_e^2 +
\rho_e^2) \left( \bigotimes_e f^{\iota_e}_{k_e\rho_e}(j_{f}) \right)
\{15j\}_{SL(2,\kompleksni)}\left( (2j_{f},2\gamma j_{f});(k_e,\rho_e) \right)
\,,\label{sfv}
\end{equation}
where the $15j$ symbol is for the unitary representations $(k,\rho)$ of the
$SL(2,C)$ group, the universal covering group of the Lorentz group. The
$f^{\iota_e}_{k_e\rho_e}$ are the fusion coefficients, defined in detail in
\cite{elpr,fk,ConradyFreidelDrugi}.

Instead of using the spin-intertwiner basis, one can rewrite
(\ref{StateSumaIntertw}) in the coherent state basis, introduced in
\cite{LivineSpeziale}. In this basis, the state sum is given by
\begin{equation} \label{StateSuma}
Z(T) = \sum_{j} \int \prod_{e,f} d^2 \vec{n}_{ef} \prod_{f\in T^*} \dim\,j_f
\prod_{v\in T^*}
W (j_{f(v)},\vec{n}_{e(v)f(v)})\,.
\end{equation}
The $\vec{n}_{ef}$ is a unit three-dimensional vector associated to the triangle dual to a face $f$
of the tetrahedron dual to an edge $e$ which belongs to $f$ (see \cite{LivineSpeziale} for details).
For a geometric tetrahedron, the four vectors $\vec n$ can be identified with the unit normal vectors for the triangles. Note that the domain of integration for each such vector is a $2$-sphere.

The key property of $W(j,\vec n)$ amplitude, which was used to find the large-spin asymptotics, is that it can be written as an integral over the manifold $SL(2,\mathbb{C})^4 \times (\mathbb{CP}^1)^{10}$, see \cite{Barret2009leprl}. More precisely,
$$ W(j,\vec n) = \textrm{const.} \prod_{k= 1}^{10} \dim j_k \,\int_{SL(2,\mathbb{C})^5} \prod_{a=1}^5 dg_a \,\delta (g_5) \,\int_{(\mathbb{CP}^1)^{10}}\prod_{k=1}^{10}dz_{k}\, \Omega (g,z)\, e^{S(j,\vec n,g,z)}  \,,$$
where $\Omega$ is a slowly changing function and
$$ S(j,\vec n,g,z) = \sum_{k=1}^{10} j_k \log w_k (\vec n,g,z) = \sum_{k=1}^{10} j_k \left(\,\ln |w_k (\vec n,g,z)| + i\theta_k (\vec n,g,z)\right) \,.$$
The functions $w_k$ are complex-valued, so that $\theta_k = \textrm{arg}\,w_k + 2\pi m_k$, where $m_k$ are integers which have to be chosen such that $\log w_k$ belong to the same branch of the logarithm. 

Since $|w_k| \le 1$, it follows that $Re\,S \le 0$ and it can be shown that the large-spin asymptotics is given by
\begin{equation} W(\lambda j,\vec n) \approx \frac{\textrm{const}}{\lambda^{12}}\sum_{x^*}\, \frac{\Omega(x^*)\,e^{i\lambda \sum_k j_k \theta_k (\vec n,x^*)}}{\sqrt{\det(- H(j,\vec n,x^*))}} \,,\label{tena}\end{equation}
for $\lambda\to +\infty$,  where the sum is over the critical points $x^* = (g^*,z^*)$ satisfying 
\begin{equation}Re\, S(j,\vec n,g^*,z^*)=0\,, \quad \frac{\partial S}{\partial g_a}{\Big |}_{x^*} = 0 \,,\quad \frac{\partial S}{\partial z_k}{\Big |}_{x^*} = 0 \,,\label{critc}\end{equation} 
and $H(x)$ is the Hessian for the function $S(x)$. There are finitely many critical points, and it can be shown that the conditions (\ref{critc}) require that $j_k$ are proportional to the areas of triangles for a geometric 4-simplex, while $\vec n$ have to be the normal vectors for the triangles in a tetrahedron of a geometric 4-simplex and $g^*$ have to be the corresponding holonomies. A geometric 4-simplex has a consistent assigment of the edge-lengths, and it can be shown that
$\theta_k (\vec n , x^*)$ is proportional to the dehidral angle for a triangle in a geometric 4-simplex, so that 
$$ S_{R}^{(v)} = \sum_{k=1}^{10} j_k \,\theta_k (\vec n , x^*) $$
corresponds to the Regge action for a 4-simplex.
  
The Hessian $ H(j,\vec n,x)$ is a $44 \times 44$ matrix, and 
\begin{equation} H_{\alpha \beta}(j,\vec n,x^*) = \sum_{k=1}^{10} j_k \, H^{(k)}_{\alpha\beta}(\vec n,x^*) \,,\label{hess}\end{equation}
since $S$ is a linear function of $j$. Consequently
\begin{equation} \det (- H) = \sum_{ m_1 + \cdots + m_{10} = 44}(j_{1})^{m_1} \cdots (j_{10})^{m_{10}} D_{m_1 ... m_{10}}(\vec n,x^*)\,, \label{deth}\end{equation}
is a homogeneous polinomial of degree $44$ in $j_k$ variables. One also has that $Re\,(-H)$ is a positive definite matrix.

\section{\label{sect3}A bound for the vertex amplitude}

We will now find a bound for the vertex amplitude by using the asymptotic formula (\ref{tena}) and its generalization for the case when some of the vertex spins are large and other are small. Since $\lambda S(j,\vec n,x) = S(\lambda j,\vec n,x)$ and 
$$ \lambda^{44} \det (- H(j,\vec n,x^*)) = \det (- H(\lambda j,\vec n,x^*)) \,,$$
then the formula (\ref{tena}) can be rewritten as
$$ W(j,\vec n) \approx \textrm{const}\,\prod_{k=1}^{10} \dim j_k \,\sum_{x^*}\, \frac{\Omega(x^*)\,e^{i \sum_k j_k \theta_k (\vec n,x^*)}}{\sqrt{\det(- H(j,\vec n,x^*))}} \,,$$
when $j = (j_1,...,j_{10}) \to (+\infty, ... ,+\infty) \equiv (+\infty)^{10}$, because $\prod_{k=1}^{10} \dim j_k$ scales as $\lambda^{10}$ for $\lambda$ large. Therefore
\begin{equation} \lim_{j \to (+\infty)^{10}}W(j,n) = \textrm{const}\,\lim_{j \to (+\infty)^{10}}
\prod_{k=1}^{10} \dim j_k \,\sum_{x^*}\, \frac{\Omega(x^*)\,e^{i \sum_k j_k \theta_k (\vec n,x^*)}}{\sqrt{\det(- H(j,\vec n,x^*))}} 
 \,. \label{ea}\end{equation}

Note that
\begin{equation} {\Big |}\sum_{x^*}\, \frac{\Omega(x^*)\,e^{i \sum_k j_k \theta_k (\vec n,x^*)}}{\sqrt{\det(- H(j,\vec n,x^*))}}{\Big |} \le \sum_{x^*}\, \frac{|\Omega(x^*)|}{\sqrt{|\det(- H(j,\vec n,g^*))}|} \,, \label{eb}\end{equation}
and
\begin{equation} \lim_{j\to (+\infty)^{10}} \frac{\prod_{k=1}^{10} \dim j_k }{\sqrt{|\det(- H(j,\vec n,x^*))}|} = 0 \,, \label{ec}\end{equation}
due to (\ref{deth}). The equations (\ref{ea}),(\ref{eb}) and (\ref{ec}) imply
\begin{equation} \lim_{j \to (+\infty)^{10}}W(j,\vec n) = 0
 \,. \label{limw}\end{equation}

The equation (\ref{limw}) is equivalent to
$$\forall\epsilon > 0 \,,\, \exists \,\delta > 0 \quad\textrm{such that}\quad j_1 > \delta \,,\cdots , \,j_{10} > \delta \, \Rightarrow |W(j,\vec n)| < \epsilon \,.$$
This implies that $W$ is a bounded function in the region 
$$D_{10} = \{ j\, |\, j_1 > \delta \,,\cdots , \,j_{10} > \delta \} \,.$$ 
If we denote with $D_m$ the region where $m<10$ spins are greater than $\delta$ and the rest are smaller or equal than $\delta$, then
$$ \mathbb{R}_+^{10}\setminus D_{10} = \bigcup_{m=0}^9 D_m \,.$$
Since the regions $D_m$ are not compact for $m>0$, we do not know whether $W$ is bounded in these regions. In order to determine this we need to know the asymptotics of $W$ for the cases when some of the spins are large and other are small. This asymptotics can be obtained by using the same method as in the case when all the vertex spins are large.

Let $m$ be the number of large spins ($m \ge 3$ due to the triangle inequalities for the vertex spins) and let $j'=(j_1,...,j_m)$. Then
$$ S(\lambda j', j'', n,x) = \sum_{k=1}^m \lambda j'_k (\ln |w_k| + i\theta_k ) + \sum_{k=m+1}^{10}j''_k (\ln |w_k| + i\theta_k ) = \lambda S_m (j',n,x) + O(1) \,. $$
Therefore the asymptotic properties of $W(j',j'',n)$ will be determined by the critical points of $S_m (j',n,x)$. Consequently
\begin{equation} W(\lambda j' , j'' , \vec n) \approx \frac{\textrm{const}}{\lambda^{r/2 - m}}\sum_{x^*}\, \frac{\Omega(x^*)\,e^{i\lambda \sum_{k=1}^m j'_k \theta_k (\vec n,x^*)}}{\sqrt{\det(- \tilde H_m (j',\vec n,x^*))}} \,, \label{mja}\end{equation}
where $r$ is the rank of the Hessian matrix $ H_m$ for $S_m$ at a critical point $x^*$ ($ 1 \le r \le 44$) and $\tilde H_m$ is the reduced Hessian matrix. $\tilde H_m$ is the restriction of the Hessian $H_m$ to the orthogonal complement of $Ker\, H_m$ and $\tilde H_m$ has to be used if $r < 44$.

The asymptotics (\ref{mja}) implies that the function $W(j',j'',\vec n)$ will vanish for large $j'$ if $r/2 - m > 0$. If this was true for all $m$ we could say that $W(j)$ is a bounded function in $\mathbb{R}_+^{10}$. However, calculating the values for $r$ is not easy. Instead, we are going to estimate $|W(j',j'',\vec n)|$. Note that (\ref{mja}) is equivalent to
$$ W( j' , j'' , \vec n) \approx \textrm{const}\,\prod_{k=1}^m \dim j'_k \,\sum_{x^*}\, \frac{\Omega(x^*)\,e^{i\lambda \sum_{k=1}^m j'_k \theta_k (\vec n,x^*)}}{\sqrt{\det(- \tilde H_m (j',\vec n,x^*))}}  $$
for $j' \to (+\infty)^m$, since $S_m$ and $\tilde H_m$ are linear functions of the spins $j'$ and $\det (-\tilde H_m)$ scales as $\lambda^r$, while $\prod_{k=1}^m \dim j_k$ scales as $\lambda^m$ when $j' \to \lambda j'$ and $\lambda$ is large. Hence
$$ \frac{W( j' , j'' , \vec n)}{\prod_{k=1}^m \dim j'_k} \approx \textrm{const}\,\sum_{x^*}\, \frac{\Omega(x^*)\,e^{i\lambda \sum_{k=1}^m j'_k \theta_k (\vec n,x^*)}}{\sqrt{\det(- \tilde H_m (j',\vec n,x^*))}} \,, $$
for $j' \to (+\infty)^m$.

From here it follows that for every $m \ge 3$
$$ \lim_{j\to (+\infty)^m } \frac{W( j' , j'' , \vec n)}{\prod_{k=1}^m \dim j'_k} = 0\,, $$
since $r(m) \ge 1$. Given that $W=0$ in $D_1$ and $D_2$, it follows that
$W(j,\vec n )/\prod_{k=1}^{10} \dim j_k$ is a bounded function in $\mathbb{R}_+^{10}$. Therefore exists $C> 0$ such that
\begin{equation}  \frac{|W( j , \vec n)|}{\prod_{k=1}^{10} \dim j_k} \le C \,. \label{wod}\end{equation}
This bound can be rewritten as
\begin{equation} |W(j,\vec n)| \leq C \prod_{k=1}^{10} \dim j_k \,,\label{psb}\end{equation}
which is convenient for investigating the absolute convergence of the state sum.

\section{\label{FinitenessSection}Finiteness}

We showed in the previous section that the vertex amplitude divided by the product of the dimensions of the vertex spins is a bounded function of spins. This result suggests to introduce a rescaled vertex amplitude $W_p$ as
\begin{equation} \label{Verteks}
W_{p} (j_f, \vec{n}_{ef} ) = \frac{W (j_f ,\vec{n}_{ef})}{\prod_{f=1}^{10} (\textrm{dim}\,j_f)^p } \, ,
\end{equation}
where $p \ge 0$, in order to improve the convergence of the state sum.

Given a triangulation $T$ of a compact
four-manifold $M$, we will consider the following state sum
\begin{equation} \label{nStateSuma}
Z_p  = \sum_{j_f} \int \prod_{e,f} d^2 \vec{n}_{ef} \prod_{f\in T^*} \textrm{dim}\,j_f
\prod_{v\in T^*}
W_{p} (j_{f(v)},\vec{n}_{e(v)f(v)})\,.
\end{equation}
It is sufficient to consider $T$ without a boundary, since if $Z(T)$
is finite, then $Z(\Gamma, T)$ will be finite due to gluing properties, where
$\Gamma$ is the boundary spin network.  

The convergence of $Z_p$ will be
determined by the large-spin asymptotics of the vertex amplitude $W$ and the values of $p$. Since the asymptotics of $W$ is not known completely, we will use the estimate (\ref{psb}) in order to find the values of $p$ which make the state sum $Z_p$ convergent.

Since
\begin{equation} \label{ZpreAproksimiranjaVerteksa}
|Z_p| \leq \sum_{j_f} \int \prod_{e,f} d^2 \vec{n}_{ef} \prod_{f\in T^*} \textrm{dim}\,j_f
\prod_{v\in T^*}
\frac{| W (j_{f(v)},\vec{n}_{e(v)f(v)}) |}{\prod_{f\in v}
(\textrm{dim}\,j_{f(v)})^p} \,,
\end{equation}
and by using (\ref{psb}) we obtain
$$
|Z_p | \leq C^V \sum_{j_f} \int \prod_{e,f} d^2\vec{n}_{ef} \prod_{f\in T^*} \textrm{dim}\,j_f
\prod_{v\in T^*}
\frac{1}{\prod_{f\in v} (\textrm{dim}\,j_{f(v)})^{p-1}}\,,
$$
where $V$ is the total number of vertices in the triangulation $T$. At this
point the integrand does not depend anymore on $\vec{n}_{ef}$, so the
appropriate integration over $4E$ $2$-spheres can be performed. Here $E$ is the
total number of edges in $\sigma$, and it is multiplied by $4$ since every edge
is a boundary for exactly four faces. After the integration we obtain
\begin{equation}
|Z_p | \leq C^V (4\pi)^{4E} \sum_{j_f} \prod_{f\in T^*} \textrm{dim}\,j_f
\prod_{v\in T^*}
\frac{1}{\prod_{f\in v} (\textrm{dim}\,j_{f(v)})^{p-1}} \, .\label{ests}
\end{equation}

The sum over the spins in (\ref{ests}) can be rewritten as a product of single-spin
sums. Let $N_f$ be the number of vertices bounding
a given face $f$. Each vertex contributes with a factor $(\dim j_f)^{-p+1}$, so
the total contribution for each face $f$ is $(\dim j_f)^{1 - (p-1) N_f}$. Thus
we can rewrite (\ref{ests}) as
\begin{equation}
|Z_p| \leq C^V (4\pi)^{4E} \prod_{f\in T^*} \sum_{j_f\in
\frac{\prirodni_0}{2}} (\dim j_f )^{1- (p-1) N_f} \, .\label{sins}
\end{equation}

The sum in (\ref{sins}) will be convergent if
$$ 1 - (p-1)  N_f < -1 \,,$$ 
or
\begin{equation} \label{ConvergenceCondition}
p-1 > \frac{2}{N_f} 
\end{equation}
for every $N_f$. Since $N_f \geq 2$ for every face $f$, a
sufficient condition for $p$ is
\begin{equation} \label{UslovNaKoeficijentP}
p > 2  \,.
\end{equation}

Therefore $Z_p$ is absolutely convergent for $p>2$, which means that it is convergent for $p > 2$. As far as the convergence of $Z_p$ for $p\le 2$ cases is concerned, one has to calculate the ranks of the Hessians $H_m$ and use the following inequalities
\begin{equation} |\det (-\tilde H_m )| \ge C_m \left(\prod_{k=1}^{m} \dim j_k \right)^{r/m}\, ,\label{hmb}\end{equation} 
when possible. We expect that the inequalities (\ref{hmb}) will hold for all $m$, since $\det(-\tilde H_m)$ is a homogeneous polinomial of the spins of the degree $r$ and $Re\,(-\tilde H_m )$ is a positive definite matrix. Then
\begin{equation} |W(j,\vec n)| \leq C_q \left(\prod_{k=1}^{10} \dim j_k \right)^{1-q}\,,\label{ib}\end{equation}
for any $j$, where $q = \textrm{min} \{ r/2m \,|\, m = 3,...,10\}$. Since $q > 0$, the new bound (\ref{ib}) will be an improvment of the bound (\ref{psb}) and consequently $Z_p$ will be absolutely convergent for
\begin{equation} p > 2 - q  \,. \label{bp}\end{equation} 
Given that $r=44$ for $m=10$, this implies that $q \ge 1/18$ ($r=1$ and $m=9$ case) and therefore $p > 35/18$.

\section{\label{ConclusionSection}Conclusions}

We proved that the deformed partition function $Z_p$ for the EPRL-FK spin foam model is convergent for $p>2$. 
We expect that the bound for $p$ can be lowered below $2$, since the inequalities (\ref{hmb}) are likely to be true. In this way one can obtain that $p> 35/18$ without calculating the matrices $ H_m$.

In order to find the exact value for $q$, the ranks $r$ of the Hessians $H_m$ have to be calculated. If it turns out that $q>2$, then the formula (\ref{bp}) will give that the $p=0$ case is convergent. However, if it turns out that $q\le 2$, then the convergence of the $p=0$ case has to be checked by some other method.

If the $p=0$ state sum is finite, our construction provides an infinite number of new models with better convergence properties. In any case, one has to decide which choices for $p$  are physical. This can be done by analyzing the semiclassical limit of the corresponding EPRL/FK model. As shown in
\cite{MVeffact1,MVeffact2}, the parameter $p$ appears in the first-order
quantum correction to the classical Einstein-Hilbert term. It is therefore an
experimental question to determine the value of $p$, provided that quantum gravity is described by an EPRL/FK spin foam model.

Given that a $p$-deformed spin foam model is finite for $p> 2$ and any choice of the triangulation $T$, one can construct a quantum field
theory whose Feynman diagrams are in one-to-one correspondence with the transition amplitudes for all
triangulations $T$, see \cite{qft,grev}. Since
all those amplitudes are finite by construction, the corresponding quantum field theory will be
perturbatively finite. For such a theory, no regularization scheme is necessary
and there is no necessity for a perturbative renormalization procedure.

As the final remark, note that  
\begin{equation}
Z(T) = \sum_{T'\subset T} Z'(T')\,,
\label{ssf}
\end{equation}
where $T'$ is a sub-complex of $T$ obtained by removing one or more
faces from $T$ and $Z'$ is the state sum where the zero spins are absent.
The state sums $Z'$ are considered more physical, because their spin foams
correspond to simplicial complex geometries where all the triangles have a
non-zero area. The relation (\ref{ssf}) was used in \cite{RovelliSmerlak2010} to
define the sum over the spin foams, since if one chooses a very large $\sigma$,
then (\ref{ssf}) implies that $Z(\sigma)$ is the result of a sum of the physical
transition amplitudes for various spin foams. Since $Z(\sigma)$ can be made
finite for EPRL/FK model if one modifies the vertex amplitude as (\ref{Verteks}),
one arrives at a concrete realization of the idea of summing over spin foams.

\medskip

\begin{acknowledgments}
We would like to thank John Barrett for discussions. AM was partially supported by the FCT grants {\tt PTDC/MAT/69635/2006} and {\tt
PTDC/MAT/099880/2008}. MV was supported by grant {\tt SFRH/BPD/46376/2008} and
partially by {\tt PTDC/MAT/099880/2008}.
\end{acknowledgments}

\end{document}